# Group dynamic behavior and psychometric profiles as substantial driver for pedestrian dynamics

Michael Schultz[1], Lars Rößger[2], Hartmut Fricke[1], and Bernhard Schlag[2]

[1]Department of Air Transport Technology and Logistics,
[2]Department of Traffic and Transportation Psychology
Faculty of Transport and Traffic Sciences "Friedrich List"
Technische Universität Dresden
schultz@ifl.tu-dresden.de

**Abstract.** Our current research lays emphasis on the extended pedestrian perception and copes with both the dynamic group behavior and the individual evaluation of situations, and hence, rather focuses on the tactical level of movement behavior. Whereas common movement models primary consider operational aspects (spatial exclusion or distance and direction related repulsion), the consideration of psychophysical concepts and intra-group coordination overcomes the idea of directed repulsion forces and derives specific movement decision with respect to the individual evaluation of situations. To provide a solid basis we analyze both data recorded at a mass event and data from a double-staged evacuation test to derive essential group dynamic behaviors and psychological related decision principles, respectively.

## 1      Introduction

Models for pedestrian dynamics cope with different aspects of behavior related to human movements. Generally, such models can be assigned to three different levels of movement characteristics: operational, tactical and strategic behavior. The basic microscopic models (e.g. social force, cellular automat, or discrete choice) particularly focus on the operational movement level. Especially, the favorable social force approach (Helbing and Molnar, 1995), which states attraction and repulsion forces between the human beings, turns out as a good analogy to reproduce substantial self-organization effects. Several model modifications and extensions of the social force model have been recently developed by the scientific community. One can notice that sustainable concepts (e.g. discretization, floor fields) will be transferred between the different model approaches, and that the models will converge as an evolutionary consequence. The definition of individual movements as destination driven processes allows for investigations of the superior self-organization effects but it neglects significant group dynamic effects (e.g. intra-group coordination) and psychological influences on (movement) decision processes. The idea of considering the individual human perception to cope with enhanced patterns of movement behavior is from our point of view the next challenge for the upcoming research tasks. An interdisciplinary research approach including the research areas of traffic sciences, sociology, mathematics, physics and psychology will ensure that the proposed methods follows a common agreement of all parties involved. In this paper we present first findings regarding the group related behavior and psychometric profiles of pedestrian and their influence to the movement dynamics.

## 2      Group Dynamics

For the data acquisition in the field, we recorded the movement behavior of the participants of the German Protestant Kirchentag at Dresden (1.-5. June 2011 with 120,000 fulltime participants and approx. 50,000 guests) and use this data as a solid base for the group constellation and behavior. As our data points out, there are significant differences in the density-speed-relation (fundamental diagram) regarding the constellation of groups. Heterogeneous crowds consists of independent pedestrians possess a homogenous density and each pedestrian has a high flexibility to change the speed and the direction of motion. The effect of *clustered density* (alternating local density clusters and open space) increases with the amount of groups, their mobility, and with the group size (fig. 1). These density spots significantly change the individual speed characteristic and the corresponding movement behavior (e.g. distance keeping, collision avoidance).

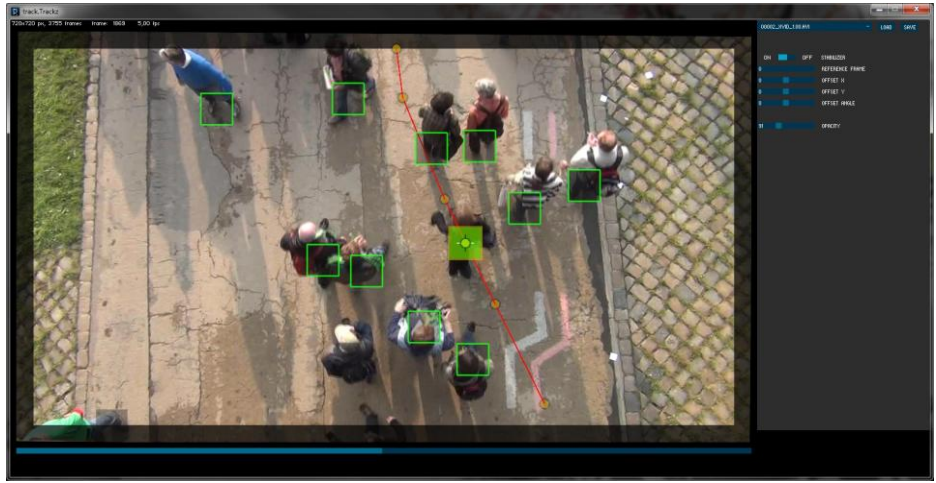

**Fig. 1.** Simple heading change of one pedestrian without the need of speed corrections in a region of intermediate density

Highly agile groups are benefit from the density clusters by efficiently use of the corresponding free accessible space. So it seemed that a mixture of two different flows exits inside the pedestrian stream. These structures are stable in an environment with lower density. With increasing density the tactical movements are barely manageable and the two separated flows are combined to one (fig. 2).

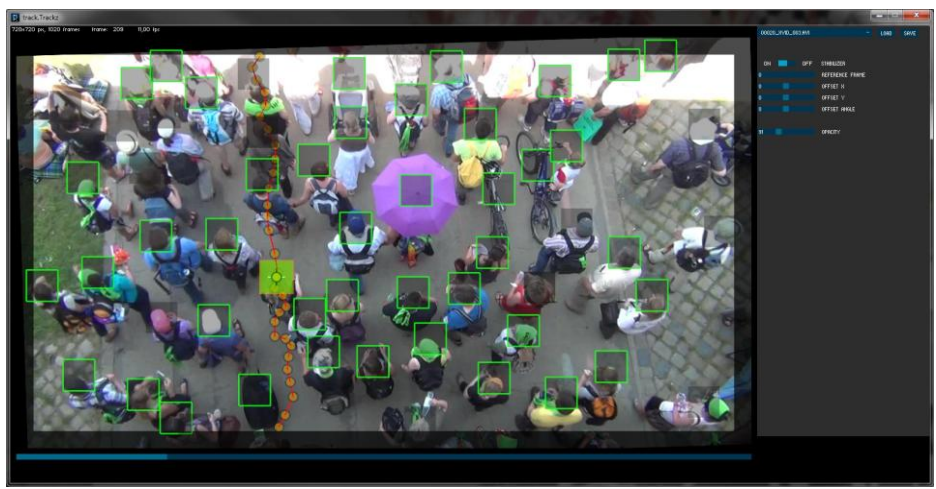

**Fig. 2.** Pedestrian movement with restricted freedom to maneuver within the crowd and starting transition from individual to coupled movement behavior

Each group member plans his individual path through the crowd and anticipates the group movements respecting the overall group constellation. Our observations suggest the assumption of an altruistically behavior as a common group movement agreement, where the group members tend to optimize the group benefit by means of trouble-free walking paths. In the course of this behavior, group members yield precedence to other members if they expect an advantage for the group. This planning procedure often based on non-verbal communication and fails with increasing density when the direct connectivity between the group members disrupts. Groups with a clear leader and follower structure as well as experienced groups contain of members with a comparable hierarchy status will efficiently solve this synchronized movement task even under crowed situations, which is confirmed by our comprehensive observations at the Kirchentag. The specific solutions to ensure an individual non-conflictive walking path results in a higher variance of speed inside the pedestrian flow compared to the expected standard distributions considering a homogenous flow.

In the following section we will provide the first results of our analyses. We strongly suppose that the ongoing data evaluation will provide further facts to validate our findings about the group behavior.

### 2.1 Environment for Identification of Groups

The main events of the German Protestant Kirchentag took place at an open space near the Elbe River framed by two bridges (Augustusbrücke and Carolabrücke, see fig. 3). Due to the fact, that the participants have to underpass these bridges to get access to the events, we could conveniently record the movement behavior of the pedestrians from a top view perspective. As agreed upon with the organizer we got access to the operational control center and could efficiently coordinate our data records with the event schedule.

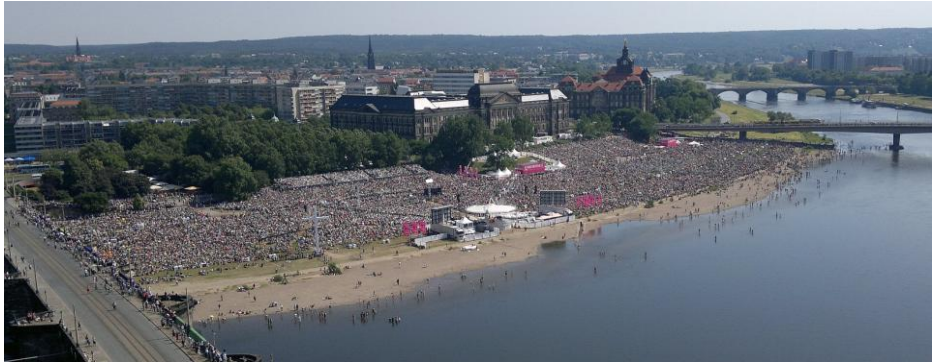

**Fig. 3.** Event area for the main mass event at German Protestant Kirchentag

For the data evaluation we developed an adequate software environment with the capability of both automatic and user defined tracking support. In particular the identification of groups the experience of human pattern recognition are not captured by common algorithms. Groups are often consisting of pedestrians in close vicinity to each other, and comparable behaviors regarding to heading, speed and acceleration. We will define this constellation as an aggregation not a group. Our approach for group detection focus on observable social behaviors as holding hands, waiting for members, or communications processes and appearance (e.g. outfit). So, the manual recognition of groups is inevitable. The process of extract pedestrian trajectories consists of 5 steps:

1. Determine the environmental characteristics (position of record device, distance to area of interest, reference point measurements) and technical specifications of the record device (field of view, distortion, zoom)
2. Calibrate the record device using reference scenarios and an independent measuring system (manual position determination in marked walking area)
3. Identification of the pedestrians and their position changes at each video frame
4. Assignment of group members (in order to distinguish individuals and groups from pedestrian aggregations)
5. Manual cross check of the pedestrian trajectories and the group assignment

### 2.2 Results

In low density environments the analysis steps for extracting the movement trajectories and the group assignment could be automatically done without user interaction, because the group members possess a clear detectable behavior: a) close distance to other members and b) prolonged separation from other groups (fig. 4). The different intra- and inter-group behavior results in a horizontal stripe formation where groups members move side by side in close distance.

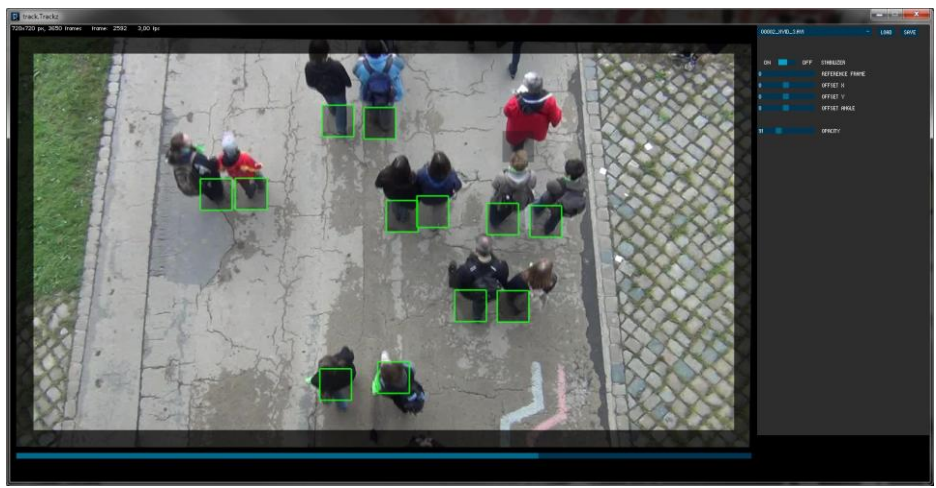

**Fig. 4.** Common group constellation, the horizontal stripe formation is a standard motion behavior at lower densities

If the separation between the groups is granted during the observation period an algorithm-based tracking provide reliable results. But aggregations of pedestrians caused by local congestions limits the quality of the group assessment. In these instances aggregations could consists of several groups which are not easily detectable by algorithms.

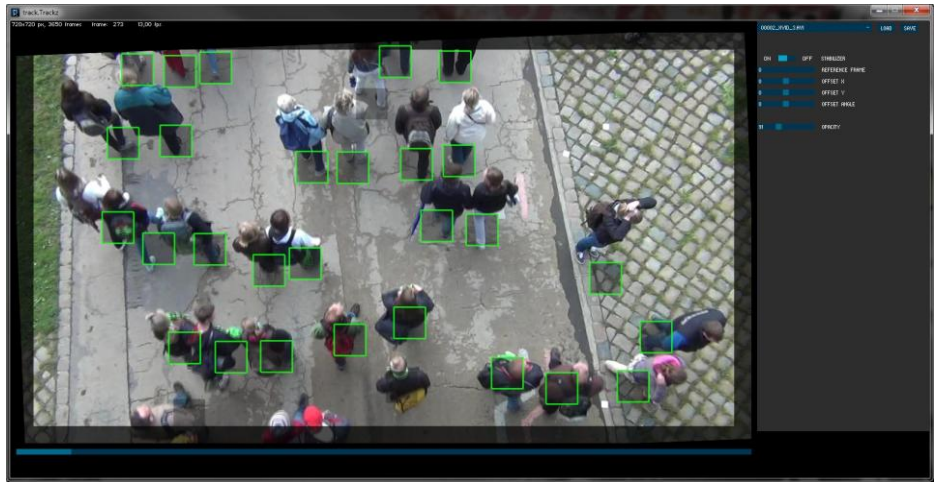

**Fig. 5.** Formations of groups and aggregation of groups/individuals demands for a combined approach of automatic algorithms and precise user intervention

The empirical knowledge of a human being allows for a significant higher assignment precision and a manual validation has to be performed to ensure valid group assignments. Nonetheless, in some situation even a manual assignment fails, so these pedestrians are not taken into account for the following statistic evaluations.

The population points out a significant emphasis on groups, only 14% of the population are single individuals. The highest proportion exhibits the group with 2 members followed by the group with 3 members. In contrast to the common population recorded by Moussaïd et al. (2010), a significant difference in the group structure can be observed (e.g. single individuals possess a rate higher than 45%).

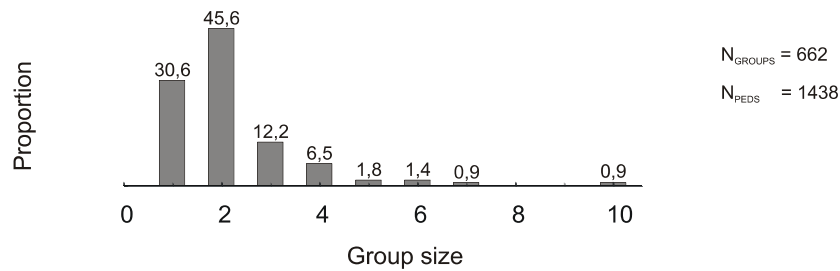

**Fig. 6.** Proportion in the observed population

The population of the German Protestant Kirchentag at Dresden (participants of a mass event) naturally differs from the populations of popular commercial walkways. But, in line with expectations the average speed is decreasing with the increasing amount of groups. As we already point out, groups tend to stay together but if the groups reached a size greater than 4 members, they intend to split up in subgroups to solve complex movement tasks (Schultz et al.,2008; Schultz, 2010). This behavior can be verified by analyzing the speed characteristic of the identified groups. To characterize the speed behavior a five-number summary descriptive statistics is used (boxplot), defined by lower and upper boundaries as well as the 25%, 75% quartile (Q.25, Q.75) and the median (50% Quartile).

The lower and upper boundaries are calculated using the 1.5 multiple of the interquartile range (IQR = Q.75 - Q.25), whereas the lower and upper values are within this range. While the median and the standard deviation of the walking speed of the groups continuously decrease with the group size, this trend interrupts at a group size of 5 with an increase of the median speed value. In addition to this interrupt, the standard deviation of the speed increases as well at groups with more than 5 members (attention should be paid to the small number of occurrences within these group categories: 5 members (12), 6 members (9), 7 members (6) and a group with 10 members occurs 6 times). To derive statistically valid results this first analysis has to be continued.

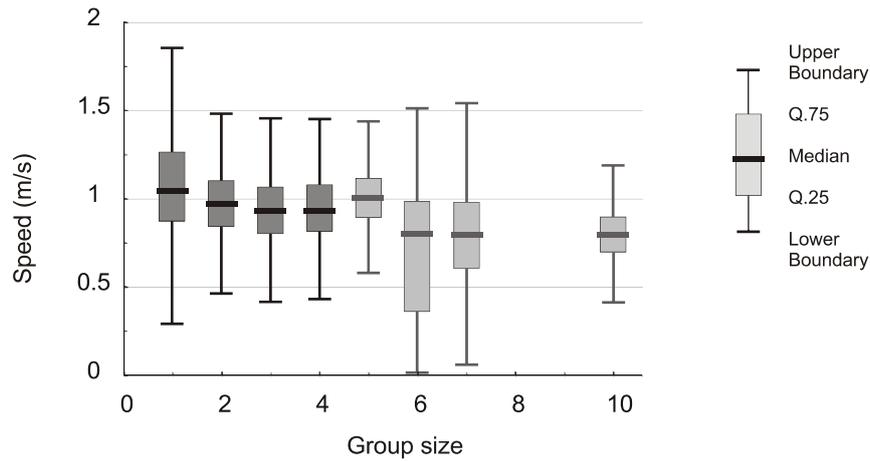

**Fig. 7.** Five-number summary descriptive statistics of speed distribution

The speed characteristics are a common procedure to determine the group behavior. To analyze the nature of the prior identified horizontal stripe formation in detail, the groups are scanned for appropriate candidates, demanding that all group members move free of external interference. For the analysis of groups with 2, 3, and 4 members 87, 19 and 10 candidates are identified, respectively. The position of each candidate at each recorded video frame (25 Hz) is plotted and a probability density function (PDF) is calculated for the longitudinal (x-position) and transversal (y-position) component referring to the group center (for members: $center_X = 1/n\ \Sigma x_i$, $center_Y = 1/n\ \Sigma y_i$). The following figure 8 represents standardized pictures of the relative positions of group members considering the mentioned constellations. In the center of the figures a 2 x 2 m grid shows the position and the PDF of the transversal component, surrounded by the PDF's for the longitudinal positions. The PDF's are smoothed by a centered average covering ± 5 values (spatial discretization 0.01 m) and they allow for a clear separation of the group members.

The group with 2 members points out a symmetric behavior in x and y direction, where the peaks of the lateral positions have distance of 0.66 m. The group with three members shows a pattern where the pedestrian in center take slightly downstream position and the outside surrounding pedestrian take slightly upstream position. The lateral distances of the y-position peak regarding to the center pedestrian are 0.61 m to the left and 0.65 m the right. The up-/downstream pattern also occurs at the group with 4 members, where the outer members repeatedly take the upstream position. The lateral distance of the inner members is 0.66 m and the distance to their associated outside attendant is 0.54 m in both cases.

These first encouraging results of the group dynamic behavior points out the stringent necessity to focus this research area, to continue the data analysis, and to develop a model which is able to cover these important group effects.

Group with 2 members:

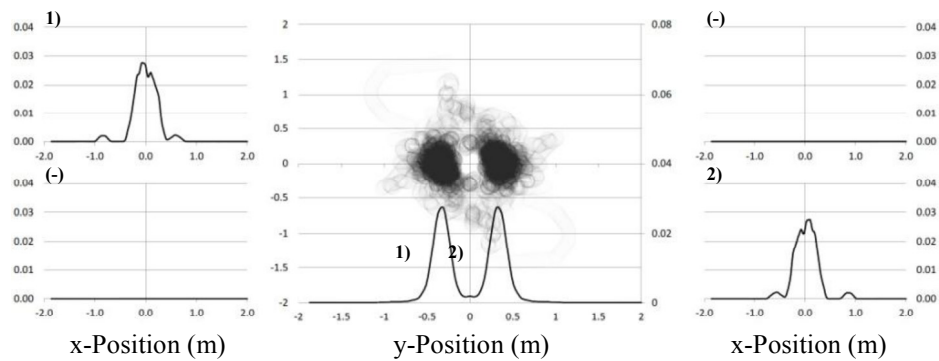

Group with 3 members:

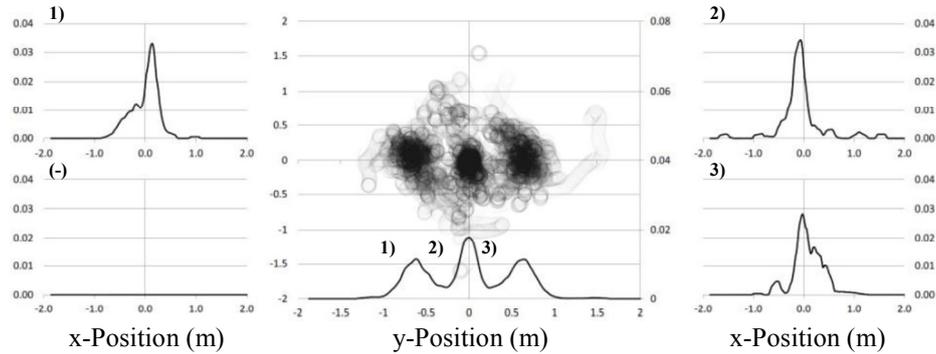

Group with 4 members:

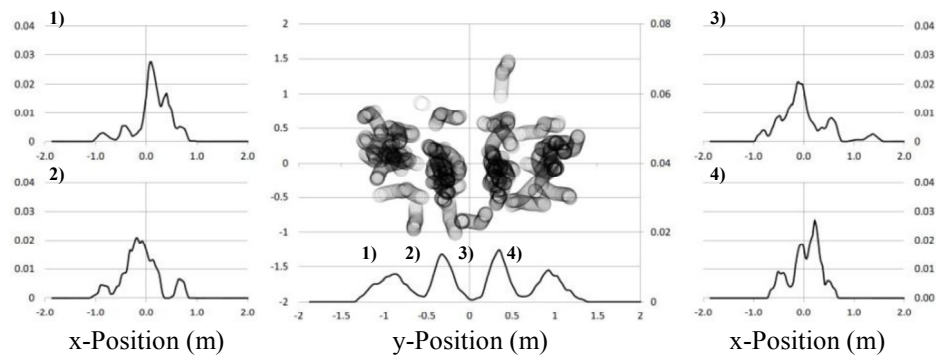

**Fig. 8.** Position of group members with ID 1, 2, 3, and 4 regarding their local distance to the group center and PDF of the respective distance component

## 3    Psychometric Profile

Our results reported above clearly indicate the existence of a *clustered density*, i.e. density spots within pedestrian crowds with effects on both individual speed characteristics and collision avoidance behavior. When focusing on the group dynamics and/or on the heterogeneity of movement pattern in pedestrian crowds the question arises of how groups emerge according to what characteristics on individual level and which influencing factors on individual level determine the heterogeneity in movement parameters? We initially proposed three different levels in order to describe movement characteristics of pedestrians: the operational level, the tactical level and the strategic level. The consideration of movement related decisions on the tactical level depending on individual factors seem to be crucial to address this question. Human beings' decision-making processes are described as bounded-rational (e.g. Simon, 1959), that is, decisions are subjected to constraints e.g. referring to the information intake, mental workload and cognitive processing. Thus, the usage of heuristics within decision-making provide valuable strategies in order to cope with cognitive boundaries and many authors (e.g. Korbus, Proctor & Holste, 2001; Carvalho, dos Santos & Vidal, 2005) claimed that decision outcomes are rather satisfying than optimizing. In that course, decision strategies are considerably affected by individual states such as (psycho-) physiological activation (e.g Harrison & Horne, 2000), affective load (for review see Loewenstein & Lerner, 2002), respectively, moods (e.g. Kuvaas & Kaufmann, 2004). Individuals' states as responses to certain situational characteristics are in turn linked to personality traits of the subjects. In the context of this paper, we focus on the role of personality traits for movement related decisions and movement parameters in an evacuation situation which is characterized by uncertainty and time pressure. Such scenarios particularly represent demands on decision making processes due to: a.) the fact that often decision rules are not available caused by the uniqueness of the situation (no familiarity), b.) threatening nature and its implications on affective work load and physiological activation, and c.) the time pressure which requires a fast decision.

When should I start to go for the exit: immediately or do I have at least enough time in order to collect my belongings and leave then for exit? Which exit should I take: the nearest one or maybe the one which is afterwards closer to the bus station? In the event of an evacuation, decisions on such questions have to be made and one might assume that they will be made without in-depth consideration about pros and cons for each alternative. The use of heuristic strategies enables us to react also in situations when deliberated thinking and/or thoughtful reasoning about alternatives becomes critical. One strategy might be to look what are doing the others around me. The *herding* effect represents the outcome of such a strategy. E.g., Baddeley et al. (2010) considered herding as a time-saving decision-making heuristic and hypothesized that certain personality types will be more likely to use herding heuristic as a decision-making shortcut. In their research on decisions in the financial context the authors found that herding is less likely amongst extraverted and emphatic individuals. The construct extraversion is one of three dimensions of personality according to the personality type model Eysenck (1968) proposed. Extraverted individuals are described as highly interested in social activities and social contacts, as acting spontaneous, are impulsive and tending to risky behavior. Furthermore, it is claimed that extraverts tend to show fast and frequent motor reactions whereas introverts stronger persist in perceptual activities and the analysis of sensory information. In the context of movement pattern in evacuation scenarios, one might assume that extraversion is negatively associated with decision time (start time) for evacuation because of their tendency to fast reactions and positively associated with the movement time because parts of the information processing is shifted into the phase of movement execution. So, for extraverts we hypothesize a more reactive movement pattern characterized by stronger variances in walking speed whereas introverts should exhibit a more anticipative movement pattern with less variance in walking speed. With respect to the question about who will be the leader of group, we assume that extraverts will start earlier than others, and thus, might provide the social cue for others to follow.

Neuroticism as further personality factor in Eysenck's model refers to the emotional stability of individuals. Individuals scoring high on neuroticism are described as emotionally labile and as tending to show difficulties to recover after emotional experiences. Moreover, neuroticism is associated with a general pattern of faster response times when detecting a threat stimulus or identifying a target cued by a threat stimulus. Decision-making processes can be interfered by dysfunctional thoughts about threating issues, and therefore, might become suboptimal with respect to the best achievable options. It is difficult to formulate assumption about the effect of neuroticism on movement related decision and movement pattern. One might assume that neurotic individuals show stronger reactions than other to the signal for evacuation, corresponding with being faster prepared to evacuate or faster walking speed toward the exit. On the other hand, it might be plausible that neurotic individuals exhibit more likely than others uncertainty and less self-confidence about what ought to be done, resulting in delays for own decisions and a stronger orientation to the behavior of others. Thus, one might also hypothesize that neurotic individuals are more prone to herding behavior than others. In order to answer the questions and/or to examine our hypotheses we conducted an explorative study on the effects of personality traits on movement related decisions and movement pattern in evacuation scenarios. To our knowledge, it is the first approach to address these issues.

## 3.1 Experimental Environment

To determine the effect of psychometric profiles data from an evacuation experiment at Technische Universität Dresden are used. One week in advance to this evacuation practice, we conducted a questionnaire survey with students in the same lecture hall and the same student course (lesson 1). The questionnaires contained an ID for the student filled in. The same student course was again asked for participating in a questionnaire survey one week later (lesson 2). The same ID was used, and additionally, the questionnaire contained an ID for the seat number within the lecture hall. Both surveys took no longer than 20 min at beginning of a lesson. After two third of total time of the second lesson, the building has been evacuated. Announcement was given by loud speakers (via speech and auditive warning signals), and the whole building was evacuated. No specific reason about the evacuation background was told to students. Location of seats and students movements were recorded by 360 degree video camera mounted at the ceiling of the lecture hall (fig. 9).

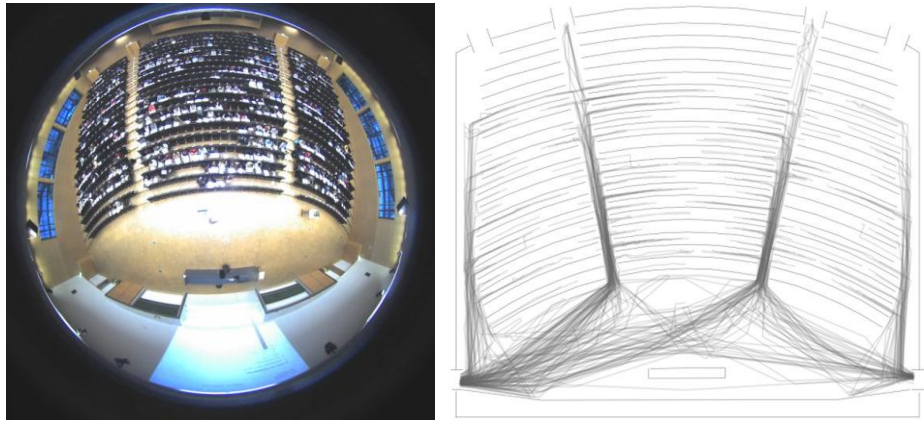

**Fig. 9.** Data acquisition with a 360 degree video camera taken at lecture hall and extraction of movement trajectories mapped to a virtual environment

At the beginning of lesson 1, we used the German version of the EPQ-RK (Ruch, 1999) in order to obtain the relevant personality traits. During lesson 2, we used a questionnaire focusing on anxiety and stress. Results presented here exclusively base on the data originated from the EPQ-RK. The EPQ-RK is a standardized instrument measuring extraversion, neuroticism and psychoticism with 50 items in total. Movement parameters of students were analyzed with video tracking and visualization software (fig. 10).

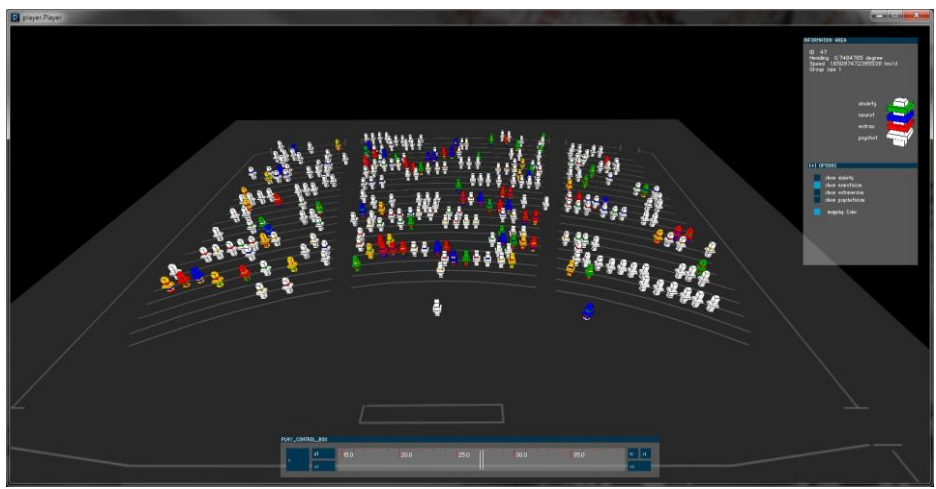

**Fig. 10.** Virtual environment for display egress progress and psychometric profiles

Data on movements and data on personality traits have been matched by linking the IDs from both questionnaires (lesson 1 and 2) and the seat numbers contained in the questionnaire for the second lesson. A number of N = 301 students took part in lesson 2 (which was evacuated in the end) and 86 % (N = 259) of the students in this lesson filled in the questionnaire. 54.4% (N = 141) were male, and 45.6% were female students. The mean age was 21.8 years (SD = 1.96) and ranged between 19 and 31 years. Participants were absolutely naive about the focus of the questionnaire survey and were told the purpose was to consider personality and stress during the stadium. From this sample we could assign 188 questionnaires (72%) to the movement data set. From this sample of N = 188 students from lesson 2 we could identify N = 88 students also in lesson 1 (46%), and thus, complete their dataset with the personality data.

### 3.2 Results

For the pre-movement phase we considered the time point individuals were ready to evacuate the lecture hall ($t_{ready}$) and the time point individuals actually left their initial locations for the exit ($t_{start}$). The difference between these two parameters to the group mean values($t_{DEVready}$, $t_{DEVstart}$) was computed for each individual and served as dependent variables. The group mean value was defined as the mean value per seat row and lecture hall sector, whereby a group consisted at least of three individuals. All values were z-standardized (fig. 11). That is, individuals with a value $t_{DEVready} < 0$ were faster prepared to start for the exit than the others seated in same row whereas $t_{DEVready} > 0$ indicated that the individual needed longer time compared to the other group members. Likewise, for $t_{DEVstart}$ less than zero the individual started earlier for the exit than the others and vice versa. According to their extraversion scores and their neuroticism scores individuals were assigned to low score (E⁻, N⁻) and high score groups (E⁺, N⁺) per median split. Pre-movement parameters were subjected to t-test analysis for independent samples to compare group norm deviations for E⁻ and E⁺, and respectively, for N⁻ and N⁺.

We considered walking speed, speed deviation and a detour index as variables within the movement phase. In addition to the personality factors neuroticism and extraversion, the walk distance was examined as a further factor. We assumed that speed deviation, mean speed as well as detour were considerably affected by the distance to walk, and were interested in potential interaction effects between distance and personality factors on the movement variables. Results indicated a significant difference for $t_{DEVready}$ between individuals with E⁻ (M = -0.28, SD = 0.92) and individuals with E⁺ (MW = 0.23, SD = 0.89); t(61) = -2.127, p = 0.037. Furthermore, the analysis revealed a significant difference for $t_{DEVstart}$ between the E⁻ sample (M = -0.42, SD = 0.90) and the E⁺ sample (M = 0.21, SD = 0.87); t(61) = -2.728, p = 0.008. Contrary to our hypothesis, these findings suggest that introvert more likely belong to the persons who started earlier within a given group whereas extravert were more likely reluctant and started with delay to the group norm. Spearman rho analysis confirmed a moderate positive correlation, corresponding with $r_s(61) = 0.40$, p = 0.001 between the absolute extraversion scores and $t_{DEVstart}$. Concerning neuroticism, t-tests did neither show significant differences between N⁻ and N⁺ for $t_{DEVready}$; t(63) = 1.574, p = 0.121, nor did for $t_{DEVstart}$; t(63) = 1.556, p = 0.125. Spearman rho analysis indicated a weak, negative correlation of $r_s(63) = -0.232$; p = 0.063 between neuroticism scores and $t_{DEVstart}$ which was marginally significant (p = 0.063). According to the latter, individuals who are more emotionally unstable tend to go for the exit earlier than others within a group.

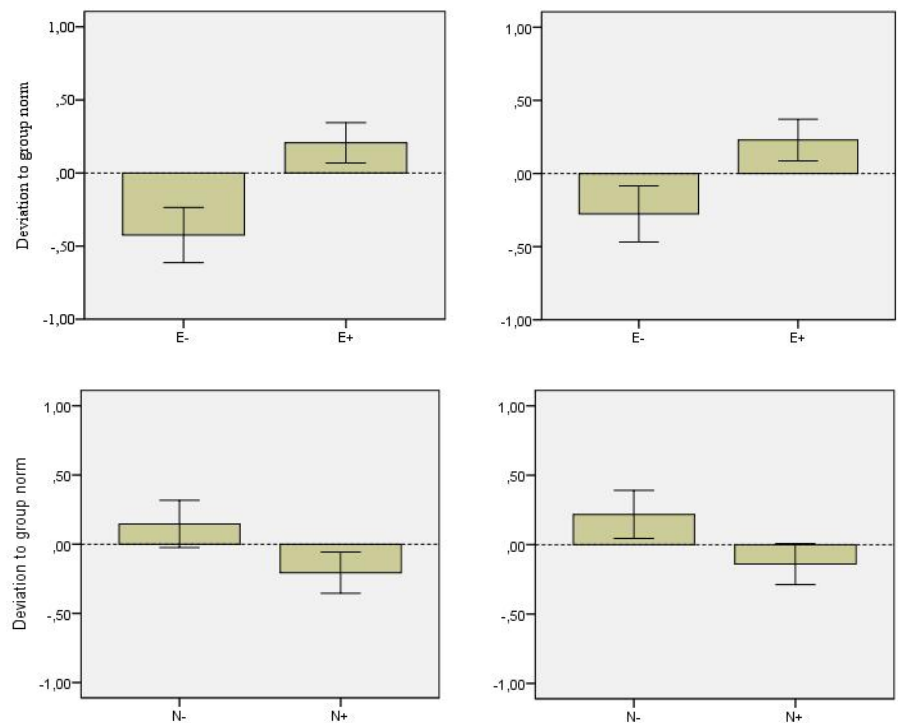

**Fig. 11.** Mean deviations to the group mean (and standard error) in time: *Ready* to start (a) and *Time* to start (b) depending on Extraversion (above) (z-standardized); Mean deviations to the group mean (and standard error) in time *Ready* to start (a) and *Time* to start (b) depending on Neuroticism (below) (z-standardized).

Thus, walking speed, speed deviation and detour index were subjected to three-factorial ANOVAs with distance, extraversion and neuroticism as factors. Individuals were therefore assigned to two groups according to their walking distance per median split (Distance low D⁻, Distance high D⁺). The ANOVA revealed neither significant effects of distance for personality factors nor for distance on mean walking speed. For speed deviation, ANOVA identified a significant main effect for walk distance on speed deviation; F(2,75) = 4.591, p = 0.035. Moreover, we found that speed deviation tend to be higher in the N⁻ group than in the N⁺ group; F(1,75) = 3.653, p = 0.060, indicating that individuals with lower scores on neuroticism tend to show more variance in their walking speed than individuals who are more emotionally unstable. The same picture emerged when considering effects on detour index. Both walking distance, F(1,75) = 24.464, p≤ 0.001, eta² = 0.24, and neuroticism, F (1,75) = 6.556, p = 0.012, eta² = 0.08, showed significant effects on the detour index, corresponding with higher detour values with greater distances, and respectively, higher detour values for individuals who score low on neuroticism. Beyond we identified an interaction effect between walking distance and neuroticism, F(1,75) = 6.935, p = 0.01, eta$^2$ = 0.085, indicating that the differences between N$^-$ and N$^+$ in the detour index was particularly strong for long distances.

# 4   Discussion and Outlook

The present study focused on both group dynamic behavior and effects of personality traits on movement related decisions and movement pattern during an evacuation scenario. We initially assumed that extraversion has an effect on decision time in advance to actual movement to the exit, corresponding with faster decision times for extraverted individuals compared to introverted individuals. Moreover, we hypothesized that extraverted individuals would show stronger variances in their walking speed because partly the information processing will shifted from the pre-movement phase to the movement phase resulting in a more reactive movement pattern contrary to an more anticipative movement pattern (and less variance in walking speed) for introverted individuals.

Contrary to our assumptions, the results suggested that extraverted started later for the exit than introverted within a given group. One possible explanation for this effect is that social information and social evaluation by others is more important for extraverted individuals and provides the necessary feedback for the appropriateness of own behavioral pattern. In that context, social cues provided by the immediate surrounding might play a greater role for extraverted individuals, and might lead in turn, to stronger delays for reactions in social situations characterized by high degrees of uncertainty. Neuroticism was considered as a second personality factor assumed to have impacts on evacuation dynamics on individual level. Results showed that individuals high scoring on neuroticism tended to start earlier than other group members for the exit but the most striking result was that they walked for the exit in a more direct manner (less likely detour). This pattern was particularly true for long walking distances. These finding might reflect the stronger perception of urgency induced by the threat stimuli for neurotic individuals.

We are fully aware about potential caveats of our study. So, it was only feasible to examine movement pattern in relation to personality for part of the sample. There is clearly the need for further investigations considering complete samples of individuals within walking under critical circumstances. One might assume that effect reported above become even stronger. However, the results emphasize the importance of personality traits on movement behavior, even in situations of uncertainty and time pressure. Therefore, the examination presented here can be seen as first attempt in order to gain deeper insights into the relation of personality and movement decisions on tactical level and on overt movement behavior on operational level. We are furthermore fully aware that further research is needed to replicate findings presented here and also to uncover mechanism beyond.